\documentclass[aip,reprint,amsmath,amssymb]{revtex4-1}

\usepackage{graphicx}
\usepackage{dcolumn}
\usepackage{bm}

\draft 

\begin{document}


\title{Time-resolved photoemission apparatus achieving sub-20-meV energy resolution and high stability} 



\author{Y.~Ishida}
\email[]{ishiday@issp.u-tokyo.ac.jp}
\affiliation{ISSP, University of Tokyo, Kashiwa-no-ha, Kashiwa, Chiba 277-8581, Japan}
\affiliation{RIKEN SPring-8 Center, Sayo, Sayo, Hyogo 679-5148, Japan}

\author{T.~Togashi}
\affiliation{RIKEN SPring-8 Center, Sayo, Sayo, Hyogo 679-5148, Japan}

\author{K.~Yamamoto}
\affiliation{RIKEN SPring-8 Center, Sayo, Sayo, Hyogo 679-5148, Japan}

\author{M.~Tanaka}
\affiliation{ISSP, University of Tokyo, Kashiwa-no-ha, Kashiwa, Chiba 277-8581, Japan}

\author{T.~Kiss}
\affiliation{ISSP, University of Tokyo, Kashiwa-no-ha, Kashiwa, Chiba 277-8581, Japan}

\author{T.~Otsu}
\affiliation{ISSP, University of Tokyo, Kashiwa-no-ha, Kashiwa, Chiba 277-8581, Japan}

\author{Y.~Kobayashi}
\affiliation{ISSP, University of Tokyo, Kashiwa-no-ha, Kashiwa, Chiba 277-8581, Japan}
\affiliation{CREST, Japan Science and Technology Agency, Tokyo 102-0075, Japan}

\author{S.~Shin}
\affiliation{ISSP, University of Tokyo, Kashiwa-no-ha, Kashiwa, 
Chiba 277-8581, Japan}
\affiliation{RIKEN SPring-8 Center, Sayo, Sayo, Hyogo 679-5148, Japan}
\affiliation{CREST, Japan Science and Technology Agency, Tokyo 102-0075, Japan}


\begin{abstract}
The paper describes a time- and angle-resolved photoemission apparatus consisting of a hemispherical analyzer and a pulsed laser source. We demonstrate 1.48-eV pump and 5.90-eV probe measurements at the $\geq$10.5-meV and $\geq$240-fs resolutions by use of fairly monochromatic 170-fs pulses delivered from a regeneratively amplified Ti:sapphire laser system operating typically at 250~kHz. The apparatus is capable to resolve the optically filled superconducting peak in the unoccupied states of a cuprate superconductor, Bi$_2$Sr$_2$CaCu$_2$O$_{8+\delta}$. A dataset recorded on Bi(111) surface is also presented. Technical descriptions include the followings: A simple procedure to fine-tune the spatio-temporal overlap of the pump-and-probe beams and their diameters; achieving a long-term stability of the system that enables a normalization-free dataset acquisition; changing the repetition rate by utilizing acoustic optical modulator and frequency-division circuit. 
\end{abstract}


\maketitle

\section{Introduction}
\label{Intro}

Angle-resolved photoemission spectroscopy (ARPES) is a powerful method to investigate the electronic structures of matter. In ARPES, monochromatic light is shined on a crystal. Photoelectrons emitted therefrom has a pattern in kinetic energy and emission angle that replicates the electronic band dispersions. 
Continuous development is being pursued in ARPES driven by the demands for investigating the fine electronic structures of complex materials such as superconducting copper oxides and heavy-fermion compounds~\cite{RMP_ARPES}. Electron analyzers, both time-of-flight and hemispherical deflection types, have improved their energy-and-angular resolutions and throughputs by virtue of the multi-channel detection and precise control in the electron-lens ($e$-lens)~\cite{Scienta_MultiChannel94, Wannberg09, ScientaTOF13}. Light source for photoemission has also been developing considerably. There is a trend to utilize lasers, as they can deliver bright, coherent, polarized, monochromatic, or ultra-short stroboscopic light. After the pioneering implementations of pulsed lasers into photoemission spectroscopy~\cite{Haight_InP, Haight_RSI88, Haight_RSI94, Haight_SurfSciRep}, investigations of ultrafast and out-of-equilibrium phenomena became possible in the femtoseconds~\cite{PetekOgawa, Bauer_PRL01, Perfetti_TaS2, Perfetti_Bi2212, Schmitt, Bovensiepen_Rev, Gedik_Science} and into the attoseconds~\cite{Cavalieri, PhotoDelay, Krausz_Rev}. Meanwhile, 70-$\mu$eV energy resolution was achieved by utilizing the monochromaticity of the lasers~\cite{Okazaki}. 

In an ultrafast pump-and-probe method employing a pulsed laser source, a laser pulse is split into two. One pulse (pump) initiates dynamics in the sample and the other pulse (probe) snapshots the non-equilibrated state. By varying the difference in the optical-path lengths of the pump and probe pulses, the recovery from the impact can be followed in the time domain. In the case for pump-and-probe-type time-resolved photoemission spectroscopy (TRPES), the photon energy of the probe pulse has to exceed the work function of the sample typically being $\sim$5~eV. 

In TRPES, there are two methods, besides using free-electron lasers~\cite{PRL10_TaS2_Flash, NewJPhys_FLASH12}, for generating probe pulses in the deep-to-extreme ultraviolet region; namely, high-harmonic generation in gaseous media~\cite{Corkum, Kulander, Heinzmann_RSI01, Bauer_Rev} and frequency up-conversion by using nonlinear crystals~\cite{Lisowski04, Carpene_RSI09, Perfetti_RSI12, Lanzara_TrPES_RSI12}. The former has the capability to generate high-energy pulses, which opens access to full valence band~\cite{Miaja-Avila}, shallow core levels~\cite{Heinzmann_GaAs,Ishizaka_TaS2}, and wide momentum ($k$) space~\cite{TiS2e_Nature11}. Achieving better time resolution is also straightforward in this method~\cite{Krausz_Rev}, because the pulse duration shortens for higher harmonics. However, a drawback is in the energy resolution: Due to the uncertainty principle, monochromaticity and short duration of a pulse are incompatible. 90-meV energy resolution is demonstrated by using the 35.6-eV high harmonics at the overall time resolution of 125~fs~\cite{RSI13_HHG_Frietsch}. The latter method is limited to generating pulses up to $\sim$6~eV, because the nonlinear crystals lose transparency at higher energies. Nonetheless, it is suited for measurements aiming at high energy resolution and high signal-to-noise ratio (S/N) in the dataset. The latter owes to the fact that the probing pulses can be up-converted from moderately strong $\lesssim$1-$\mu$J pulses, and hence, at a high repetition rate of $>$100~kHz. Sub-30-meV time-resolved ARPES (TARPES) studies have been accessing into the carrier dynamics across the Fermi level ($E_F$)~\cite{Lanzara_TrARPES_NPhys,Lanzara_Science,Sobota_PRL12}, while sub-20-meV measurements also became feasible~\cite{Ishida_HOPG, KimPRL}. 

This paper describes the specification of a TARPES apparatus consisting of a hemispherical analyzer and a pulsed laser source. The design concept was to achieve a high energy resolution capable for detecting ultrafast responses of the fine electronic structures of complex materials. To this end, we adopted a laser source delivering fairly monochromatic, and hence, not too short pulses of 170-fs duration at a typical repetition rate of 250~kHz; namely, a regeneratively amplified Ti:sapphire laser system. For the generation of the probe pulses, we used nonlinear crystals. 

We also describe technical points for the sake of TARPES to be a precise, convenient, and hence, a versatile method for materials investigation. First of all, compared to ARPES, TARPES has another dimension of time in the dataset. In addition, the flux of the pulsed probe has to be reduced in order to minimize the space-charge broadening effect~\cite{PasslackAeschlimann_SpaceCharge,Zhou_SpaceMirror}. These characteristics of TARPES make the acquisition time of the dataset longer than that in ARPES, in general. It was therefore a mandatory to achieve a long-term stability of the apparatus in order to obtain a high S/N dataset recorded under the reduced probing flux. Second, two beams, pump and probe, have to be manipulated in TARPES, which is contrasted to the single-beam manipulation in ARPES. Moreover, the manipulation has to be done in both space and time at the focal point of the $e$-lens in the ultra-high vacuum. We sought for simple methods that allow precise manipulation and characterization of the pump and probe beams. 

The paper is organized as follows. After the present introduction (Sec.~\ref{Intro}), we describe the TARPES setup in Sec.~\ref{Setup}. In this section, we first sketch the TARPES system (\ref{layout}), and then describe the procedures to improve the time resolution (\ref{TimeResolution}) and to align the pump and probe beams to the focal point of the $e$-lens by use of some devices (\ref{align}). A method to cross-correlate the pump and probe pulses in space and time is explicated in \ref{Graphite}, in which we use pump-and-probe photoemission signal of graphite~\cite{Ishida_HOPG}. Description on the space-charge effect and stability of the TARPES system  are presented in \ref{SpaceCharge} and \ref{Stability}, respectively. In \ref{DelayScan}, we describe a unique data acquisition procedure in TARPES that becomes effective after achieving the long-term stability of the system. An optional device that varies the repetition rate is explicated in \ref{RepRate}. In Sec.~\ref{Performance}, we describe the performance of the TARPES system by presenting the dataset recorded on a cuprate superconductor in a high-energy resolution mode (\ref{Bi2212})  and on Bi(111) surface during a prolonged acquisition (\ref{Bi}). Finally, we conclude the paper in Sec.~\ref{Conclusions}.

\section{Experimental setup}
\label{Setup}

\subsection{TARPES layout}
\label{layout}

\begin{figure}[htb]
\begin{center}
\includegraphics[width=8cm]{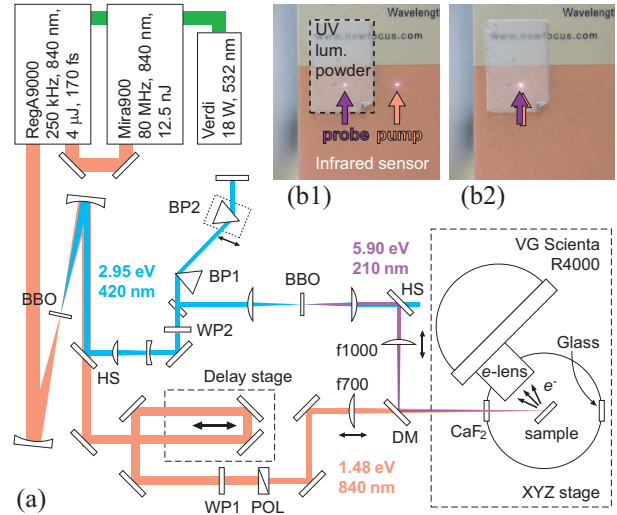}
\caption{\label{fig_setup} 
TARPES layout. (a) Schematic view of the TARPES system consisting of a pulsed laser source, pump and probe beam lines, and a hemispherical electron analyzer. The laser system consists of a Ti:sapphire regenerative amplifier (Coherent RegA 9000) seeded by a mode-locked Ti:sapphire oscillator lasing at 840~nm (Coherent Mira). Ti:sapphire crystals of the amplifier and oscillator are pumped by a Nd:YVO$_4$ laser system (Coherent Verdi18) that delivers 532-nm beam up to 18~W. (b) A sensor device simultaneously visualizing the infrared pump and ultraviolet probe beams. Snapshots taken before (b1) and after (b2) overlapping the pump and probe beams. 
}
\end{center}
\end{figure}

The general layout of the TARPES setup is sketched in
Figure~\ref{fig_setup}(a). 
A Ti:sapphire laser system (Coherent RegA9000) repetitively generates 170-fs laser pulses with center photon energy at 1.48~eV (840~nm). The repetition rate $\Omega$ is tunable in the 60\,-\,300~kHz range, and set typically to 250~kHz. Using type I non-linear second-harmonic generation at two $\beta$-BaB$_2$O$_4$ (BBO) crystals, a portion of the laser pulse is up-converted into a 5.90~eV (210~nm) probing pulse. The first and second BBOs are 0.3- and 0.1-mm thick, respectively. The pump and probe beams are split by a harmonic separator (HS) located after the first BBO. A translational delay stage varies the optical-path length of the pump beam line. The pump and probe beams are finally joined at a dichroic mirror (DM) just before entering the ultra-high vacuum chamber (base pressure $<$5\,$\times$\,10$^{-11}$~Torr) through a CaF$_2$ window. Photoelectrons are collected by a hemispherical analyzer (VG Scienta R4000) equipped with an $e$-lens and a multi-channel-plate (MCP) detector. 

The intensity of the pump beam is controlled by a combination of a polarizer (POL) and a $\lambda$/2 wave plate (WP1). By rotating the wave plate, we vary the amount of the polarization component that is passed though by the subsequent polarizer. The probe beam intensity is controlled through a similar mechanism:  A $\lambda$/2 wave plate (WP2) regulates the polarization component that is passed through by subsequent Breuster prism (BP1). The wave plates WP1 and WP2 are mounted on rotary stages that are controlled from a personal computer (PC).

\subsection{Improving the time resolution}
\label{TimeResolution}

A light pulse elongates after traveling through optical elements and air that act as dispersive media. The group-velocity dispersion (GVD) has to be either minimized or compensated in order to increase the time resolution in TARPES. GVD for a $\sim$200-fs pulse becomes pronounced at wavelength $\lambda$\,=\,200\,-\,400~nm. Particular care thus has to be taken in the probe beam line. 

We inserted a prism-pair compressor~\cite{PrismCompressor} in the probe beam line [Figure~\ref{fig_setup}(a)]: A pair of Breuster prisms, BP1 and BP2, are positioned between the first and second BBOs at the angle of minimum deviation. By installing the prism-pair compressor, the time resolution of TARPES improved from $\sim$400 to $\gtrsim$240~fs, as described later in Sec.~\ref{Graphite}. We have also kept the optical-path length of the 210~nm probe in air (from the second BBO to the CaF$_2$ window) to be less than 80~cm, because the GVD becomes pronounced when $\lambda$ approaches the absorption edge of air at $\sim$205~nm. GVD in the pump beam line was negligibly small, which was confirmed by measuring the duration of the pump pulse just before the CaF$_2$ window by a scanning-type autocorrelator (A$\cdot$P$\cdot$E Mini). 

Optimization of the prism-pair compressor is done by shifting BP2 into the beam through a two-step process. First, the amount of the BP2 insertion is determined by maximizing the photoemission count rate, which is a good measure of the up-conversion efficiency at the second BBO, and hence the shortness of the pulse. In the second step, we maximize the contrast of the pump-and-probe signal of graphite in the unoccupied side (see later in Sec.~\ref{Graphite}). This ensures that the pump duration is optimized for the TARPES measurements. The time resolution of $<$300~fs was easily achieved just by undergoing the first step.

\subsection{Alignment of the pump and probe beams}
\label{align}

\begin{figure}[htb]
\begin{center}
\includegraphics[width=5.5cm]{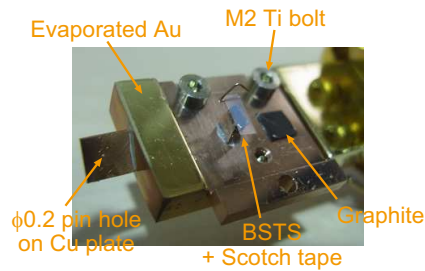}
\caption{\label{fig_Pinhole} 
Setup around the sample. A layered BSTS sample~\cite{KimPRL} to be cleaved by a Scotch-tape method is on a copper plate that is screwed to the cryogenic manipulator by M2 titan bolts. Graphite (already cleaved) is also attached next to the sample and is used as a reference. A pin hole device is also installed. $E_F$ is calibrated by recording the Fermi cutoff of evaporated Au in electrical contact to the sample and the analyzer. 
}
\end{center}
\end{figure}

In a pump-and-probe method, the pump and probe beams are usually overlapped on the sample with the pump beam size slightly larger than that of the probe. In the case for TARPES using a hemispherical analyzer, another stringent condition exists: The spatial overlap of the pump and probe beams has to be done at the focal point of the $e$-lens, which is located $\sim$35~mm from the $e$-lens aperture in the ultra-high vacuum chamber. We here describe a procedure to direct the pump and probe beams to the focal point of the $e$-lens at the desired beam sizes. The key items are a sensor device [Figure~\ref{fig_setup}(b)], an XYZ stage [Figure~\ref{fig_setup}(a)], a microscope~\cite{Muro}, and a pin-hole device (Figure~\ref{fig_Pinhole}).

First, we coarsely align the pump and probe beams co-linearly before the CaF$_2$ window in air. The alignment can be done conveniently by using a sensor device that visualizes the infrared (IR) and ultraviolet (UV) beams simultaneously. The sensor device is composed of an IR sensing card attached with UV fluorescence powder by a double-sided adhesive tape, as shown in Figure~\ref{fig_setup}(b). The UV probe glows on the surface of the tape, while the IR pump penetrates through the tape and glows on the card. Coarse focusing of the pump and probe beams is also done at this point by varying the translational stages of the f1000 and f700 lenses that focus the pump and probe beams, respectively. 

Next, we shift the focal point of the $e$-lens into the coarsely-aligned beam. This can be done because the entire analyzer chamber is on a XYZ stage that is separated from the laser table. The alignment using the XYZ stage is done by centering and sharpening the photoelectron image on the MCP detector. The use of a long-working-distance microscope employing K2/S lens of Infinity Photo-Optical Co.~\cite{Muro} is helpful to know the beam position in the analyzer chamber. The microscope is viewed by a CCD camera, which has sensitivity to the pump beam in the IR. Therefore, after the co-linear alignment in air, as described previously, the pump beam seen through the CCD on sample is also the locus of the probe beam. The advantage of using the XYZ stage is that the alignment can be done without losing the co-linearity of the pump-and-probe beam. Without the XYZ stage, pump and probe beams have to be moved one by one, during which the locus of the invisible UV probe beam is lost in the CCD image. 

We next optimize the beam sizes by utilizing a pin hole drilled on a Cu plate attached to the sample mount; see Figure~\ref{fig_Pinhole}. The advantage of the method presented below is that the tuning and estimation of the beam sizes are done at the focal point of the $e$-lens.  

The probe beam size is estimated as follows: Similarly to doing photoemission spectroscopy on a small sample, we search the pin hole of 200~$\mu$m in diameter by monitoring the photoelectron count rate, which becomes minimal when the pin hole is properly aligned to the focal point of the $e$-lens.  Finiteness in the beam size, or the tail of the Gaussian-like beam, generates photoelectrons from the copper plate surrounding the hole. From the amount of the loss of the photoelectron count rate on the 200-$\mu$m pin hole, we can estimate the beam size of the probe. Tightening or loosening of the focus on the pin hole is done by monitoring the loss of the count rate from the pin-hole device. 

After the above procedure, the following three are spatially matched; the focal point of the $e$-lens, the pin hole, and the properly-sized probing beam. Under this configuration, we finally fine-tune the alignment of the pump beam and its size. This is done by maximizing the pump-beam intensity $I$ passing through the pin hole. The intensity $I$ is monitored by a power meter set after the viewing port behind the sample; see Figure~\ref{fig_setup}. Note, the standard glass viewing port transmits the pump beam but blocks the probe beam; the latter can be seen on the glass as a florescence spot, which is also useful for visualizing the probe beam in the ultraviolet. The pump beam size is estimated from $I/I_0$, where $I_0$ is the intensity of the pump beam, or the read of the power meter when the pin-hole device is fully retracted from the optical path. By tuning the f700 focusing lens, and hence $I$, we optimize the pump beam size on the pin hole, which is identical to the focal point of the $e$-lens.

\subsection{Cross correlating the pump and probe pulses by using graphite}
\label{Graphite}

We find highly-oriented pyrolytic graphite (HOPG) as a useful cross correlator of the pump and probe pulses~\cite{Ishida_HOPG}. Calibration of the origin of the pump-and-probe delay $t_0$, fine-tuning of the spatial overlap of the pump and probe beams, and estimation of the time resolution $\varDelta t$ are conveniently done by using the pump-and-probe photoemission signal of graphite at the focal point of the $e$-lens. Graphite can be easily cleaved and the exposed surface is stable over weeks in the vacuum chamber. These characteristics also facilitate us to use graphite as a reference sample. We cleave graphite by attaching Scotch tape on surface and peel it off in the ultra-high vacuum at room temperature; see Figure~\ref{fig_Pinhole}: To our experience, Scotch tape is compatible to ultra-high vacuum unless baked. 

\begin{figure}[htb]
\begin{center}
\includegraphics[width=8.0cm]{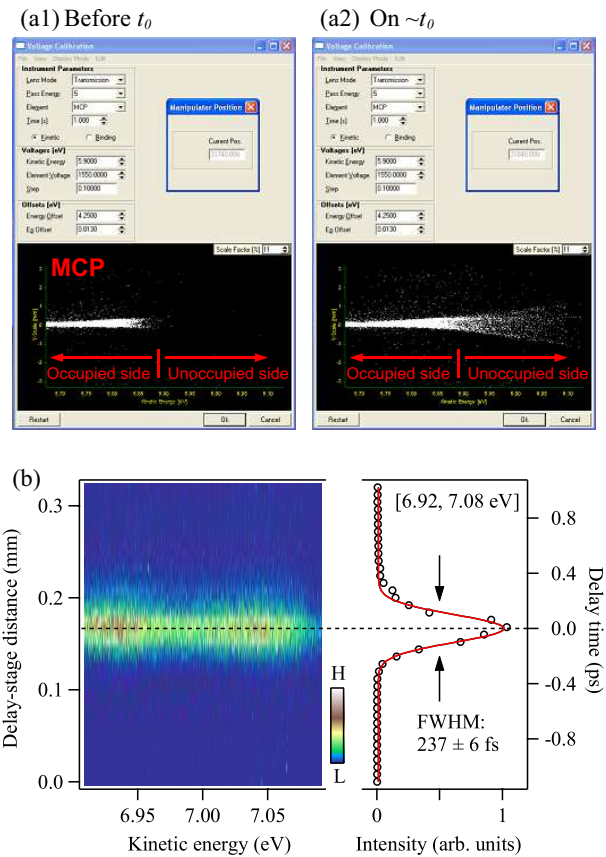}
\caption{\label{fig_HOPG} 
Pump-and-probe signal of graphite as a cross correlator. 
(a) Real-time photoemission signal displayed on the PC screen. White band in the MCP image is the photoemission signal dispersed horizontally in energy. When the spatio-temporal overlap of the pump and probe pulses is small (a1), photoelectron signal appears only in the occupied side. When the overlap becomes large (a2), the signal also appears in the unoccupied side. Fine tuning of the spatial overlap of the pump beam to the probe beam is done by maximizing the signal in the unoccupied side. 
(b) Determination of $t_0$ and time resolution $\varDelta t$. Pump-and-probe photoelectron signal around 7-eV kinetic energy, or 1.1~eV above $E_F$, is mapped during the delay-stage scan around $t_0$ (left panel). Right panel shows the integrated photoelectron intensity as a function of the delay-stage distance, or delay time. Gaussian fitting gives the locus of $t_0$ and $\varDelta t$\,=\,237\,$\pm$\,6~fs.  
}
\end{center}
\end{figure}

When pumped by a 1.48-eV pulse, graphite exhibits photoemission signal in the unoccupied side that is strong enough to be observed in real time. Figure~\ref{fig_HOPG}(a) shows the PC screens displaying the photoemission image on the MCP detector. When the temporal overlap of the pump and probe pulses are small, the photoemission signal appears only in the occupied side (a1). As the delay stage is scanned, photoemission signal in the unoccupied side appears when the temporal overlap becomes large (a2). Final fine tuning of the spatial overlap of the pump beam to the probe beam is also done at this point by maximizing the signal in the unoccupied side around $t_0$. Pulse compression by the prism pair (the second step described in Sec.~\ref{TimeResolution}) is also done by monitoring the contrast between the on-$t_0$ (a2) and off-$t_0$ (a1) images. 

The time resolution and origin of the delay are determined by using a resolution-limited fast response of graphite that appears at $\sim$1~eV above the Fermi level~\cite{Ishida_HOPG}. The left panel in Figure~\ref{fig_HOPG}(b) shows the map of the photoemission signal from graphite around 7-eV kinetic energy, or 1.1~eV above the Fermi level, during the delay-stage scan.  Signals of photoelectrons appear when the temporal overlap of the pump and probe pulses becomes large. Integrated intensity in the kinetic-energy region [6.92, 7.08~eV] is shown in the right panel in Figure~\ref{fig_HOPG}(b). By fitting the profile with a Gaussian, we obtain the time resolution and the locus of $t_0$: In the case shown in Figure~\ref{fig_HOPG}(b), full width at half maximum (FWHM) of the Gaussian is 240~fs, which is the typical time resolution after the tuning. The acquisition time was 10~min for the dataset presented in Figure~\ref{fig_HOPG}(b). The repetitive scanning of the delay stage was done (described in Sec.~\ref{DelayScan}) in a fixed kinetic-energy mode with the pass energy ($E_{\mbox{\it{pass}}}$) and slit width set to 5~eV and 0.8~mm, respectively.

\subsection{Space-charge effect and its reduction}
\label{SpaceCharge}

\begin{figure}[htb]
\begin{center}
\includegraphics[width=8cm]{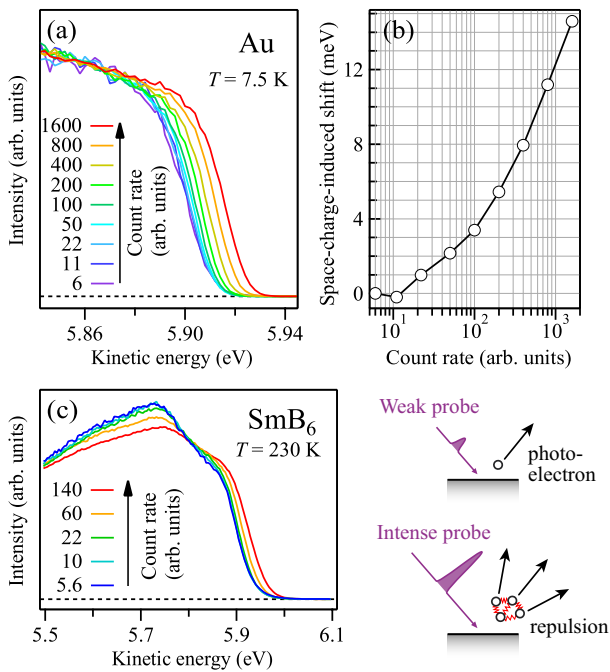}
\caption{\label{fig_SpaceCharge} 
Space-charge broadening. (a) Fermi cutoff of Au recorded at various flux of the probe beam. (b) Shift of the Fermi cutoff of Au plotted as a function of the photoemission count rate, which is a measure of the probe flux. (c) Valence-band spectra of SmB$_6$ recorded at various probe flux. The schematic figure shows that a too intense probing pulse generates a bunch of photoelectrons that repel each other through the Coulomb effect. The spectra shown in (a) and (c) are arbitrarily normalized in intensity. 
}
\end{center}
\end{figure}

In TARPES implemented by the pump-and-probe method, the flux of the probe beam cannot be increased as much as desired because of the space-charge effect~\cite{PasslackAeschlimann_SpaceCharge,Zhou_SpaceMirror}. A too intense probing pulse generates a bunch of photoelectrons that repel each other through the Coulomb effect; thereby distorting the energy-and-angular distribution of the photoelectrons; see the schematic shown in Figure~\ref{fig_SpaceCharge}. The space-charge effect has to be minimized by reducing the probe flux when fine structures such as superconducting gaps and small surface photo-voltage-induced shift of the spectra are of interest. 

Figure~\ref{fig_SpaceCharge}(a) shows the Fermi cutoff of gold (Au) recorded at various flux of the probe beam. As the flux is increased, the leading edge of the spectrum is shifted into higher kinetic energy due to the space-charge effect. The locus of the leading edge is plotted in Figure~\ref{fig_SpaceCharge}(b) as a function of the photoemission count rate, which is a good measure of the flux of the probe beam. For reducing the leading edge shift from 4~meV to $<$1~meV, the intensity of the probe has to be reduced for one order of magnitude. 

The magnitude of the space-charge effect depends on materials. Therefore, the flux of the probe has to be determined for each sample before the measurement. The case for SmB$_6$ is shown in Figure~\ref{fig_SpaceCharge}(c), in which valence-band spectra recorded at various probing flux are displayed. As the flux is increased, the leading edge is shifted into higher kinetic energy. In addition, the spectral shape is deformed. These are due to the space-charge effect, while some non-linearity of the MCP to the photoemission count rate may have some effect at high count rates. The space-charge effect is judged to be negligibly small for the spectrum recorded at the lowest flux shown in Figure~\ref{fig_SpaceCharge}(c), because the locus of the leading edge and the spectral shape hardly changed from that recorded at the second lowest flux; see the spectra recorded at the count rates of 5.6 and 10 in arbitrary units. The unchanged spectral shape also ensures that the linearity of the MCP is fairly good when the photoemission count rate is sufficiently reduced.

\subsection{Stabilizing the TARPES system}
\label{Stability}

In order to achieve good S/N in the TARPES dataset, the apparatus has to operate stably during the prolonged acquisition time. Particular care had to be taken not only in the laser power but also in the the sample position during the acquisition, as we describe below. The stability of the system was also important to realize a unique TARPES dataset acquisition scheme, which we describe in Sec.~\ref{DelayScan}. 

\begin{figure}[htb]
\begin{center}
\includegraphics[width=8cm]{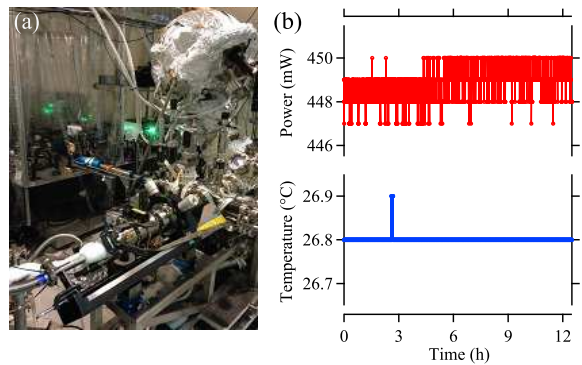}
\caption{\label{fig_stable} 
Stability of the TARPES system. (a) A snapshot during TARPES measurement at low temperatures. Mechanical supporting is done around the He return line (left bottom) where ice is developing. The vinyl sheet housing the laser source is seen in the back side. Seen in the front side is the vacuum chamber and hemispherical analyzer, which are on the XYZ stage [Figure~\ref{fig_setup}(a)]. (b) Logging of the laser power and temperature. The laser reflected from POL in the pump beam line [Figure~\ref{fig_setup}(a)] was monitored by a pyrometer-type power meter. Temperature was simultaneously monitored near the oscillator in the vinyl house. The minimal step of the reads of the digital power meter and thermometer were 1~mW and 0.1$^{\circ}$C, respectively. }
\end{center}
\end{figure}

The laser system is in a temperature-controlled environment [Figure~\ref{fig_stable}(a)]. It is housed by a vinyl sheet that prevents the system from being subjected to the flow of air; the vinyl house is in a room whose temperature is stabilized by an air conditioner equipped with a PID-controlled heater. 
Figure~\ref{fig_stable}(b) shows the temperature monitored on the laser table in the vinyl house. During the $\sim$12-hours acquisition, the read was stable at 26.8$^{\circ}$C, while fluctuations to 26.9$^{\circ}$C sometimes occurred. The laser intensity monitored simultaneously was virtually stable at 449~mW, as shown in Figure~\ref{fig_stable}(b). 
 
The sample position should not move during the data acquisition. This becomes of particular importance not only because of the prolonged acquisition time, but also because the focal point of the $e$-lens becomes tight when analyzing the low-kinetic-energy photoelectrons generated by the 5.90~eV probe, which is rather low in photon energy for a photoemission light source. In fact, when the working distance (distance between the sample and the aperture of the $e$-lens) is varied for 100~$\mu$m, slight changes can be seen in the spectral shape. 

Mechanical tightening of the cryostat that holds the sample is found to be effective; see Figure~\ref{fig_stable}(a). Here, not only the cryostat but also the return line of the vaporized He gas is also supported mechanically. After properly supporting the cryostat, the sample position does not move for 100~$\mu$m over at least 12~hours, as judged from the spectral shape. This is also true even when the He return line gains the weight of ice during the low-temperature measurement. Ice developing on the He return line can be seen in Figure~\ref{fig_stable}(a). 

The flow rate of liquid He should also be kept as constant as possible. We keep the pressure in the liquid He vessel to 0.5~MPa by using a pressure regulator attached to the vessel. $T$\,=\,10 and 3.5~K can be maintained stably for at least 7 and 3 days, respectively, when using a 250-L liquid He vessel. We do not pump out the He gas from the return line in order to avoid the vibration of the pump to be transmitted to the sample at the focal point of the $e$-lens.

\subsection{Data acquisition through repetitive scanning of the delay stage}
\label{DelayScan}

The delay stage driver and the ARPES data acquisition software are linked. Therefore, a TARPES dataset can be acquired in a automated scan. After achieving the long-term stability of the TARPES system (Sec.~\ref{Stability}), the following dataset-acquisition scheme became very effective: A TARPES sequence is designed such that the delay stage repetitively scans over the specified delay points. In this way, we can keep the same S/N in each TARPES image recorded at different delay points. The major advantage is that the intensity of the raw image recorded at each delay points is normalized to the acquisition time. That is, the spectral intensity recorded at different delays can be compared directly without any normalizations; see Sec.~\ref{Bi}. In general, when comparing the intensity of the ARPES data recorded at different conditions, there would be concerns such as the difference in the surface conditions, slight change in the sample position due to the thermal expansion of the cryostat, and so on. Therefore, when comparing the intensity of the two datasets recorded at different conditions, some kind of intensity normalization is usually undertaken.

\subsection{Varying the repetition rate}
\label{RepRate}

In a pump-and-probe measurement employing a mode-locked pulsed laser source, the sample is repetitively impinged by the pump pulses. When the pumped sample does not fully recover before the next pump arrives, the response accumulates to result in a so-called periodic steady state. One of the examples relevant to TRPES is a semiconductor surface exhibiting surface photo-voltage (SPV) effect~\cite{Marsi, Kamada}. SPV occurs as a result of the pump-induced variation of surface band bending developed on the edge of semiconductors. The duration of SPV usually exceeds micro-seconds, so that pump-induced shifts of the spectrum are observed at $t$\,$<$\,0 when the interval time of the pulses is shorter than the SPV duration. When investigating a periodic steady state in a pump-and-probe method, interests may appear in its interval-time dependence. 

As an option, we can insert a device that varies the interval time, or the repetition rate. This is done by using an acoustic optical modulator (AOM) drove by 80-MHz carrier frequency (CF). When CF is applied to the AOM crystal (TeO$_2$), acoustic wave is generated which acts as a grating that diffracts out the laser passing through the crystal. By using an AOM driver composed of frequency division and timing microchips, we can synchronically gate the AOM to the amplifier output at a repetition rate of $\Omega/2^n$ ($n$\,=\,0, 1, ..., 6; $n$ can be set manually). The AOM device can be inserted just after the amplifier, and the diffracted pulses are sent into the pump and probe beam lines, so that TARPES is done at the $\Omega/2^n$ repetition rate. 

The repetition rate dependency of the pump-induced shift of SmB$_6$ spectra, which we attribute to the SPV effect, is presented elsewhere~\cite{SmB6}.

\section{Performance}
\label{Performance}

\subsection{TARPES of Bi2212: Observation of the superconducting peak in the unoccupied side}
\label{Bi2212}

We present TARPES data of Bi$_2$Sr$_2$CaCu$_2$O$_{8+\delta}$ (Bi2212), a copper oxide superconductor that has a critical temperature $T_c$ of $\sim$92~K~\cite{Mochiku}. The electronic structure of Bi2212 is quasi-two dimensional and has strong anisotropy in $k$ space~\cite{RMP_ARPES}. The superconducting gap closes in the nodal direction (0,\,0)\,-\,($\pi$,\,$\pi$), since Bi2212 is a $d$-wave superconductor. In the anti-nodal direction, or around ($\pi$,\,0) in $k$ space, there is a pseudo-gap region, which is characterized by the persistence of a gap even above $T_c$~\cite{Norman_Nature98} and by a particle-hole asymmetric electronic structure around $E_F$~\cite{Hashimoto_NPhys10}. Between the nodal and anti-nodal directions, a particle-hole symmetric structure across $E_F$ is found below $T_c$, namely the Bogoliubov bands in the occupied and unoccupied sides separated by a superconducting gap~\cite{WSLee_nature}. This near-nodal region recently gained renewed interest, whether the superconducting gap disappears at $T_c$ similarly to that in the BCS theory of superconductivity~\cite{WSLee_nature} or whether it persists slightly above $T_c$, which may be in conformity to the precursor pairing picture~\cite{Dessau_NaturePhys12,Dessau_PRB}. For a recent review that includes the above discussions in the near-nodal region, see Ref.~\cite{NPhys_ARPES}.

\begin{figure}[htb]
\begin{center}
\includegraphics[width=8cm]{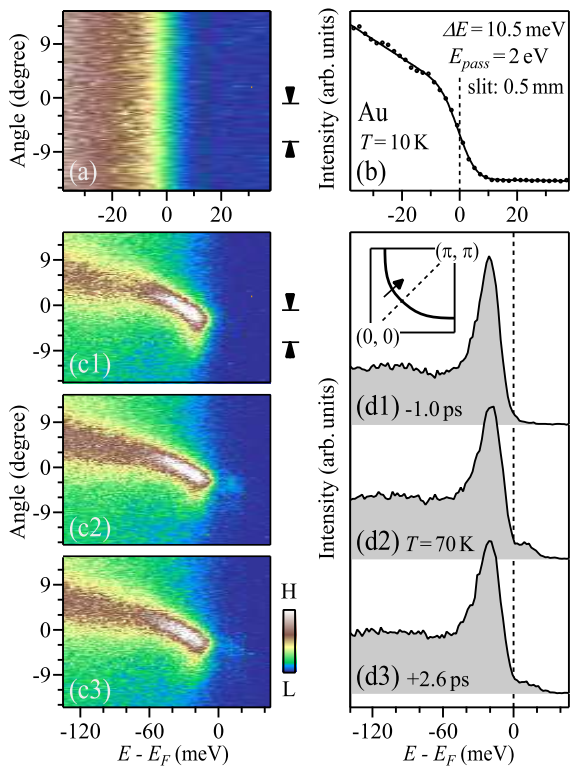}
\caption{\label{fig_Bi2212} 
TARPES of Bi2212. (a) ARPES image of Au around the Fermi cutoff recorded at 10~K. (b) Fermi edge of Au. A Gaussian convolved Fermi-Dirac distribution function is overlaid. The EDC was obtained by integrating the angular region of the ARPES image indicated in (a). (c) Quasi-particle dispersions of Bi2212 recorded before the pump at 10~K ($t$\,=\,-1.0~ps; c1), without pump at 70~K (c2), and after the pump at 10~K ($t$\,=\,2.6~ps; c3). (d) EDCs of Bi2212. EDCs of (d1)-(d3) were obtained from the corresponding images of (c1)-(c3), with the integration done in the angular region indicated in (c1).  
}
\end{center}
\end{figure}

A high energy resolution is a prerequisite to TARPES when the superconducting gaps in the near-nodal regions become of interest, because the gap size becomes diminishingly small on approach to the node. We set the TARPES spectrometer in a high-energy resolution mode (slit width of 0.5~mm and $E_{\mbox{\it{pass}}}$ of 2~eV) and investigated the pump-induced changes of the quasi-particle dispersion of Bi2212 in the near-nodal region. 

We first present in Figure~\ref{fig_Bi2212}(a) the ARPES image of Au around the Fermi cutoff recorded at $T$\,=\,10~K. The spectra were recorded prior to the Bi2212 measurement, and  were used for energy calibration. We had set the probe intensity so that the space-charge induced shift was much less than 1~meV. The corresponding energy distribution curve (EDC) is displayed in Figure~\ref{fig_Bi2212}(b), which is the integral of the ARPES image in the angular region indicated by arrows in Figure~\ref{fig_Bi2212}(a). By fitting the EDC to a Fermi-Dirac distribution function convolved with a Gaussian, the energy resolution (FWHM of the Gaussian) was estimated to be 10.5~meV. 

Figure~\ref{fig_Bi2212}(c1) displays the TARPES image of Bi2212 at $t$\,=\,-1.0~ps along a near-nodal cut in $k$ space indicated in the inset of (d). The sample was held at $T$\,=\,10~K, and the pump density per pulse was set to $p$\,=\,14~$\mu$J/cm$^{-2}$. The particle-hole symmetry across $E_F$ in this cut was confirmed by heating the sample up to 70~K and confirming the thermally populated superconducting peak in the unoccupied side; see the quasi-particle dispersion recorded at $T$\,=\,70~K in Figure~\ref{fig_Bi2212}(c2) and the corresponding EDC shown in (d2), which  exhibits a peak at $\sim$10~meV above $E_F$. The TARPES image recorded at $t\,=$\,2.6~ps is shown in Figure~\ref{fig_Bi2212}(c3). A shoulder or a peak is observed at $\sim$10~meV in the corresponding EDC [\ref{fig_Bi2212}(d3)], which is attributed to the pump-induced population of the superconducting peak in the upper Bogoliubov band. Such features were not reported in the previous TARPES studies done at the energy resolution of $>$20~meV~\cite{Perfetti_Bi2212,Lanzara_TrARPES_NPhys,Bovensiepen,Lanzara_Science,Lanzara_NCom}. We thus succeed in the direct observation of the unoccupied side of the superconducting gap by TARPES in the high energy resolution mode. 

All the Bi2212 data presented in Figure~\ref{fig_Bi2212}(c) were recorded after reducing the space-charge-induced shift to $<$1~meV, and their energy axes were calibrated by using the Fermi cutoff of Au [Figure~\ref{fig_Bi2212}(a)] as done in usual ARPES data analyses. It is warned in the literature that photoemission spectroscopy utilizing intense pulsed light source can suffer from the Fermi level referencing problem~\cite{PasslackAeschlimann_SpaceCharge, Zhou_SpaceMirror}: Extrinsic energy shift and broadening due to the space-charge and possible mirror-charge effects can occur depending on the photon flux as well as the sample configuration such as the tilting angle. The Fermi level referencing using Au presented herein is judged to be reliable, because the calibrated Fermi level nicely falls into the internal energy reference of the sample; see Figures~\ref{fig_Bi2212}(d2) and ~\ref{fig_Bi2212}(d3), in which the Fermi level is nicely located in-between the superconducting peaks in the occupied and unoccupied side.  It is thus demonstrated that the Fermi level referencing to a metal, as done in usual ARPES studies, is equally safe in TARPES, if the photon flux of the probe beam is sufficiently reduced.

\subsection{TARPES of Bi thin film grown on HOPG}
\label{Bi}

We show a dataset of TARPES recorded by scanning the delay stage repetitively; see Sec.~\ref{DelayScan}. The measurement was performed on a thin film of bismuth (Bi) grown on HOPG. Vacuum-evaporated Bi on cleaved HOPG is found to form micro-crystals with the 111 face oriented normal to surface. Because HOPG consists of layers of micron-sized crystallite sheets that have random in-plane orientation, the 111 surface of the Bi micro-crystals are also randomly oriented in the basal plane; see later. Details of the film growth and characterization will be presented elsewhere~\cite{Ishida_YbFiber}. 

\begin{figure}[htb]
\begin{center}
\includegraphics[width=8.6cm]{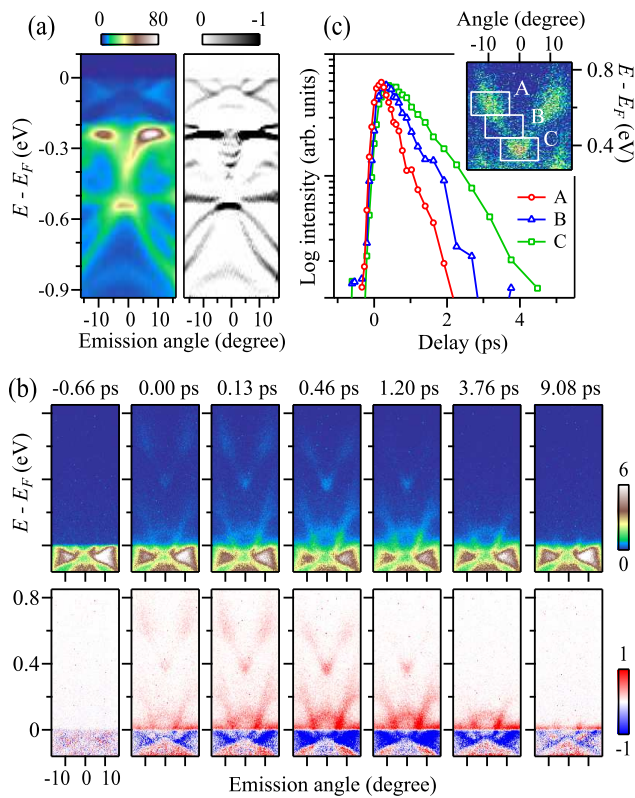}
\caption{\label{fig_Bi} 
TARPES of micro-crystalline Bi thin film grown on HOPG. (a) Band dispersion recorded at $t$\,$\leq$\,-0.6~ps, namely, before the arrival of the pump pulse (left) and its second derivative in energy (right). (b) TARPES images. Top panels show band dispersions, and bottom panels show difference to the averaged image before pumped.  (c) Intensity variations along the V-shaped band in the unoccupied side. Inset shows the magnified view of the V-shaped band recorded at $t$\,=\,0.13~ps and the integration windows A-C along the band. The intensity scales adopted in the ARPES, TARPES, and difference images presented in (a) and (b) are in common arbitrary units. 
}
\end{center}
\end{figure}
 
The left panel of Figure~\ref{fig_Bi}(a) shows an average over 12 TARPES images recorded at $t$\,$\leq$\,-0.6~ps during a sequence composed of 58 delay points. Since the probing is done before the arrival of the pump pulse, band dispersions in the occupied side are observed as in usual ARPES. Characteristic electronic structures of the 111-oriented Bi thin film are nicely resolved such as the Rashba-split surface states dispersing at $>$-0.15~eV~\cite{YuKorteev} and quantized states occurring around -0.3\,$\pm$\,0.1~eV~\cite{Hirahara_PRL06, Hirahara_PRB07}. The latter is clearly observed in the second-derivative image shown in the right panel, in which the color scale highlights the negative curvature in energy of the original image, and hence, the band dispersions. Some of the electronic structures are more clearly resolved than those reported in the literature, such as the crossing point of the Rashba-split branches at $\bar{\varGamma}$~\cite{YuKorteev, Ast}. The clear view in the ARPES images presented herein not only owes to the high energy resolution of 17~meV adopted in the measurement (slit width and $E_{\mbox{\it{pass}}}$ were set to 0.8~mm and 2~eV, respectively) but also to the magnified view in $k$ space around $\bar{\varGamma}$ when probed by the low excitation energy of~5.9 eV. Since the probe beam illuminates a large number of micron-sized Bi(111) surface that have random in-plane orientation, the band dispersions that warp hexagonally in going away from $\bar{\varGamma}$ are all projected on the TARPES images. For example, the dispersions of the Rashba-split surface states along $\bar{\varGamma}$\,-\,$\bar{M}$ and $\bar{\varGamma}$\,-\,$\bar{K}$ directions, the former contributing to a lobe-shaped Fermi surface~\cite{YuKorteev, Ast}, are simultaneously observed in the ARPES images shown in Figure~\ref{fig_Bi}(a). 

The upper and lower panels of Figure~\ref{fig_Bi}(b) respectively display selected TARPES images and their difference to the averaged image before pumped. Note, the difference was taken without any data processing such as the intensity normalization, because the intensity of the raw data is already normalized to the acquisition time; thanks to the repetitive delay-stage scanning (Sec.~\ref{DelayScan}) and stability of the TARPES system during the data acquisition of $\sim$12~hours (Sec.~\ref{Stability}). On arrival of the pump ($t$\,=\,0 and 0.13~ps), spectral intensity is spread not only into the unoccupied side of the Rashba-split surface states but also into a V-shaped band dispersing at $>$0.4~eV. After the pump-induced filling of the unoccupied states, recovery dynamics follows over 10~ps. The recovery of the spectral intensity along the V-shaped band is shown in Figure~\ref{fig_Bi}(c). Here, the integrated intensities in the frames A-C indicated in the inset to Figure~\ref{fig_Bi}(c) are plotted as functions of the delay time. The recovery of the intensity slows on approach to the bottom of the V-shaped band, which indicates that the pump-induced carriers are running down along the V-shaped band during the recovery. 

The above results not only show the efficacy of the normalization-free data-acquisition for capturing the pump-induced dynamics, but also demonstrate the capability of the apparatus for revealing the band dispersions in the unoccupied side at the sub-20-meV resolution. That is, the bands above $E_F$ can be investigated at the energy resolution comparable to those of standard ARPES done at modern synchrotron facilities. TARPES results on a topological insulator Bi$_{1.5}$Sb$_{0.5}$Te$_{1.7}$Se$_{1.3}$ (BSTS) are presented elsewhere~\cite{KimPRL}, in which the dispersion of the topological surface states traversing from the valence band to the conduction band is fully visualized at the energy resolution of 15~meV. The unoccupied side up to $\sim$1~eV above $E_F$ can be nicely revealed for some materials such as topological insulators~\cite{Sobota_PRL12, Gedik_Science}, Bi~\cite{Perfetti_Bi}, and semiconductors~\cite{Haight_InP, Azuma}; all of which may be categorized as low-carrier-density materials. On the other hand, in the case of metallic materials, the unoccupied states are less visible by TARPES, presumably because the pump pulse is efficiently reflected by the surface. As a result, the pump-induced changes are small and are confined within a narrow energy region around $E_F$ but not up to $\sim$1~eV above $E_F$, which may be the case for Bi2212 shown in Figure~\ref{fig_Bi2212}.

\section{Conclusions}
\label{Conclusions}
 
We described a 1.48-eV pump and 5.90-eV probe TARPES apparatus operating at $\geq$10.5-meV and $\geq$240-fs resolutions. Capability of the apparatus was demonstrated by presenting the datasets recorded on Bi2212 and 111-face-oriented micro-crystalline Bi thin film. 
In the TARPES of Bi2212, we resolved a pump-induced superconducting peak in the unoccupied side at the high-energy-resolution settings of 10.5~meV. In the TARPES of a Bi thin film, carrier dynamics as well as band dispersions above $E_F$ were revealed. A normalization-free dataset acquisition was demonstrated in the latter, which became possible after achieving a long-term stability of the TARPES system. Investigations of the variations in the TARPES intensities, such as pump-induced spectral-weight transfers in correlated materials and pump-induced changes in the photoemission matrix element would thus become possible without any post-processing in the dataset. 

We also described simple methods to align and characterize the pump and probe beams and to determine the time resolution by using graphite. After all, daily inspection of the beam alignment before TARPES measurements is done with a $\lesssim$15-min procedure: First, we coarsely check the pump-and-probe overlap in air by using the sensor device (Sec.~\ref{align}), and then check the spatio-temporal overlap and time resolution by using the pump-and-probe signal of graphite (Sec.~\ref{Graphite}) which is attached next to the sample to be measured (Figure~\ref{fig_Pinhole}). When concerns still remain in the beam alignment, we check it by utilizing the pin-hole device (Sec.~\ref{align}), which takes not more than 5~min. We hope that the descriptions would facilitate TARPES to be a versatile tool for investigating the electronic structures and dynamics of matter.

\begin{acknowledgments}
The authors acknowledge T.~Mochiku, S.~Nakane, and K.~Hirata for providing high-quality Bi2212 single crystals; F.~Iga and T.~Takabatake for providing high-quality SmB$_6$ single crystals; K.~Yaji for preparing a calibrated Bi evaporator; Y.~Ozawa for realizing the pinhole device; M.~Endo, and T.~Nakamura for developing the AOM driver; and T.~Saitoh for help in the TARPES measurements. This work was supported by Photon and Quantum Basic Research Coordinated Development Program from MEXT and JSPS KAKENHI, Grant Nos.~23740256 and 26800165. 
\end{acknowledgments}


%

\end{document}